\documentstyle[prl,aps,preprint,tighten,floats,epsf]{revtex}

\newbox\rotbox

\begin{document}
\draft
\setcounter{page}{0}
\def\footnoterule{\kern-3pt \hrule width\hsize \kern3pt}
\tighten
\title{INTERFERENCE FRAGMENTATION FUNCTIONS AND\\
       THE NUCLEON'S TRANSVERSITY\thanks{%
This work is supported in part by funds provided by the U.S.
Department of Energy (D.O.E.) under cooperative research
agreement \#DF-FC02-94ER40818}}

\author{R.~L.~Jaffe,\footnote{Email address: {\tt jaffe@mitlns.mit.edu}}
Xuemin~Jin,\footnote{Email address: {\tt jin@ctpa02.mit.edu}}
 and Jian~Tang\footnote{Email address: {\tt jtang@mitlns.mit.edu}}}

\address{{~}\\Center for Theoretical Physics\\
Laboratory for Nuclear Science \\
and Department of Physics \\
Massachusetts Institute of Technology\\
Cambridge, Massachusetts 02139 \\
{~}}

\date{MIT-CTP-2672, ~hep-ph/9709322 {~~~~~} September 1997}
\maketitle

\thispagestyle{empty}

\begin{abstract}

We introduce twist-two quark interference fragmentation functions  in
helicity density matrix formalism and study their physical implications. We
show how the nucleon's transversity distribution can be probed  through the
final state interaction between two mesons ($\pi^+\pi^-$, $K\overline K$,
or $\pi K$) produced in the current fragmentation region in deep inelastic
scattering on a transversely polarized nucleon. 

\end{abstract}

\vspace*{\fill}
\begin{center}
Submitted to {\it Physical Review Letters}
\end{center}

\narrowtext

The quark transversity distribution in the nucleon is one of the three
fundamental distributions which characterize the state of quarks
in the nucleon at leading twist. Measurements of the other two have
shed considerable light upon the  quark-gluon substructure of the nucleon.
The transversity distribution measures the probability difference to
find a quark polarized along versus opposite to  the polarization of a
nucleon polarized transversely to its direction of motion
\cite{ralston79,jaffe91,artru93,cortes92}. It is identical to the helicity
difference distribution in the non-relativistic limit where 
rotations and boosts commute. However,  we have learned from 
$g_A/g_V\ne 5/3$ and the recent measurement of the spin fraction 
carried by quarks in the nucleon, $\Sigma \approx 0.2$ \cite{spin-rev}, 
that the quarks  inside the nucleon cannot be non-relativistic.
The difference between the  transversity and helicity distributions is 
a further and more detailed measure of the relativistic nature of the 
quarks inside the nucleon. 

The transversity distribution measures the correlation of quarks with
opposite chirality in the nucleon.  Since hard scattering processes in QCD
preserve chirality at leading twist, transversity is difficult  to
measure experimentally. For example, it is suppressed like 
${\cal O}(m_q/Q)$ in totally-inclusive deep inelastic scattering (DIS). 
Ways have been suggested to measure  the transversity distribution. 
These include transversely polarized Drell-Yan \cite{ralston79}, 
twist-three pion production in DIS \cite{jaffe91,ji94},  the so-called 
``Collins effect'' as defined in single particle 
fragmentation\cite{collins94}, and polarized
$\Lambda$ production in DIS \cite{artru93,jaffe96}. However, each of 
these has drawbacks. The Drell-Yan cross section is small and requires 
both transversely polarized quarks and antiquarks in the nucleon.  
Twist-three pion production is suppressed by ${\cal O}(1/Q)$, and 
a $g_T\hat f_1$ term must be subtracted to reveal the transversity 
dependence.  The Collins effect in single particle fragmentation
requires a residual final state interaction 
phase between the observed particle and the rest of the jet
 which we believe to be unlikely (see below).  
Finally polarized $\Lambda$ production suffers from a likely low production 
rate for hyperons in the current fragmentation region and an as yet  
unknown and possibly small polarization transfer from $u$-quarks to 
the $\Lambda$.

In this Letter we develop another way to isolate the
quark  transversity distribution in the nucleon that is free from
many of these  shortcomings. We study semi-inclusive production
of two mesons ({\it e.g.\/} $\pi^+\pi^-$, $\pi K$, or 
$K\overline K$) in the current fragmentation region
in deep inelastic scattering on a transversely polarized nucleon.
The possibility to measure quark transversity distribution in the
nucleon via such a process was first suggested by Collins and
collaborators \cite{collinsetal} (see also Ref.~\cite{ji94}).
Our analysis focuses on the {\it interference} between the
$s$- and $p$-wave of the two-meson system around  the
$\rho$ (for pions), $K^*$ (for $\pi K$), or the $\phi$ (for kaons).  
We make explicit use of two-meson phase  shifts to characterize the
interference. Such an interference effect allows the  quark's
polarization information to be  carried through $\vec k_+
\times \vec k_- \cdot  \vec S_\perp$,  where
$\vec k_+$, $\vec k_-$, and $\vec S_\perp$ are the
three-momenta of 
$\pi^+$ ($K$), $\pi^-$ ($\overline K$), and the nucleon's
transverse spin, respectively. This effect is at the leading twist
level, and the production  rates for pions and kaons are large in
the current fragmentation region.  However, it would vanish by
T-invariance in the  absence of final state interactions, or by
C-invariance if the two-meson state were an eigenstate of
C-parity.   Both suppressions are evaded in the $\rho$
($\pi^+\pi^-$), $K^*$ ($\pi K$), and 
$\phi$ ($K\overline K$) mass regions.

The final state interactions of $\pi\pi$, $\pi K$, and $ K\overline{K}$ are
known in terms of  meson-meson phase shifts. From these phase shifts 
we know that $s$- and $p$-wave production channels interfere strongly 
in the mass region around the $\rho$, $K^*$, and $\phi$ meson resonances. 
Since the $s$- and $p$-waves have opposite C-parity, the interference 
provides exactly the charge conjugation mixing necessary.
Combining perturbative QCD, final state interaction
theory, and data on the meson-meson phase shifts, we can relate this 
asymmetry to known quantities, the transversity distribution we seek, and to
a new  type of fragmentation function that describes the $s$- and
$p$-wave interference  in the process
$q\rightarrow \pi^+\pi^- (\pi K, K\overline{K})$. 
Unless this fragmentation is 
anomalously small, the measurement of this asymmetry  may be the most
promising way to measure the quark transversity distribution.

Earlier works \cite{collins94} have explored
angular correlations of the form $\vec k_1\times\vec k_2\cdot\vec
S_\perp$, where $\vec k_1$ and $\vec k_2$ are vectors characterizing
the final state in DIS.  The simplest example would be $\vec k_1=\vec
q$ and $\vec k_1=\vec k_\pi$, the momentum of a pion.  These
asymmetries, however, require that the final state interaction phase
between the observed hadron(s) and the rest of the hadronic final
state must not vanish when the unobserved states are summed over.  We
believe this to be unlikely.  We utilize a final state
phase generated by the two-meson final state interaction, which is
well understood theoretically and well measured experimentally.  

The results of our analysis are summarized by Eq.~(\ref{asymmetry})
where we present the asymmetry for $\pi^+\pi^- (\pi K, K\overline{K})$
production in the current fragmentation region.  Current data on
$\pi\pi$, $\pi K$, and $K\overline{K}$ phase shifts are used to estimate the
magnitude of the effect as a function of the two-meson invariant mass
(see Fig.~\ref{fig2}).


We consider the semi-inclusive deep inelastic scattering process with 
two-pion final states being detected: $e\vec N_\bot\rightarrow 
e'\pi^+\pi^- X$. The analysis to follow applies as well to 
$\pi K$ or $K\overline{K}$ production. The nucleon 
target  is transversely polarized with polarization vector $S_\mu$. 
The electron beam is unpolarized. The four-momenta of the initial and final
electron are denoted by $k = (E, {\vec k})$ and $k^\prime = (E^\prime,  {\vec
k^\prime})$, and the nucleon's momentum is $P_\mu$. The momentum of the
virtual photon is $q=k-k^\prime$, and $Q^2 =
-q^2=-4EE'\sin^2{\theta/2}$,  where $\theta$ is the electron scattering
angle. The electron mass  will be  neglected throughout. We adopt the
standard variables in DIS, $x = Q^2/2P\cdot q$  and  $y = P\cdot q/P\cdot k$.
The $\sigma [(\pi\pi)^{I=0}_{l=0}]$ and $\rho [(\pi\pi)^{I=1}_{l=1}]$
resonances are produced in the current fragmentation  region with momentum
$P_h$ and momentum fraction $z = P_h\cdot q/q^2$. We recognize that the
$\pi\pi$ s-wave is not resonant in the vicinity of the $\rho$ and our
analysis does not depend on a resonance approximation.  For simplicity we
refer to the non-resonant $s$-wave as the ``$\sigma$''.  
In their work on the two pion system \cite{colladin94}, Collins and 
Ladinsky made the unphysical assumption of a narrow
$s$-wave resonance interfering with a (real) continuum $p$-wave, neither of
which appears in $\pi\pi$-scattering data.

The invariant squared mass of the two-pion system is 
$m^2 = (k_++k_-)^2$, with $k_+$ and $k_-$ the momentum of
$\pi^+$ and 
$\pi^-$, respectively. The decay polar angle in the rest frame of the
two-meson system is denoted by $\Theta$, and the azimuthal angle $\phi$ is
defined as the angle of the normal of two-pion plane with respect to the
polarization  vector $\vec S_\perp$ of the nucleon, 
$\cos\phi = {{\vec k_+}\times{\vec k_-}\cdot\vec S_\perp / |\vec k_+\times
\vec k_-||\vec S_\perp|}$. This is the analog of the ``Collins angle'' defined 
by the $\pi^+\pi^-$ system \cite{collins94}.

To simplify our analysis we make a collinear approximation, i.e., 
$\theta\approx 0$, in referring the fragmentation coordinate system to
the axis defined by the incident electron (the complete analysis will be 
published elsewhere\cite{JJT2}).  At SLAC, HERMES, and COMPASS
energies, a typical value for $\theta$ is less than $0.1$.  
Complexities in the analysis of
fragmentation turn out to be proportional to $\sin^2\theta$ and can be 
ignored at fixed target facilities of interest. In this approximation the
production of two pions can be viewed as a collinear process with the
electron beam defining the common $\hat{e}_3$ axis. Also we take
$\vec S_\perp$ along the $\hat{e}_1$ axis.

\begin{figure}[h]
\begin{minipage}[h]{6.0in}
\vspace*{1cm}
\epsfxsize=7.0truecm
\centerline{\epsffile{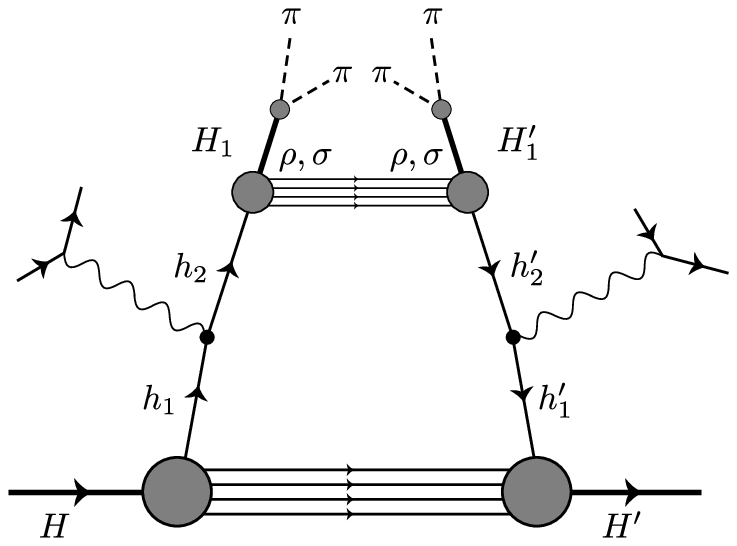}}
\vspace*{1cm}
        \caption{{\sf Hard scattering diagram for $\pi^+\pi^-
(K\overline{K})$ production in
the current fragmentation region of electron scattering from a target
nucleon. In perturbative QCD the diagram (from bottom to top) factors
into the products of distribution function, hard scattering,
fragmentation function, and final state interaction. Helicity density matrix
labels are shown explicitly.}}
        \label{fig1}
\end{minipage}
\end{figure}

\begin{figure}[h]
\begin{minipage}[h]{6.0in}
\epsfxsize=7.0truecm
\centerline{\epsffile{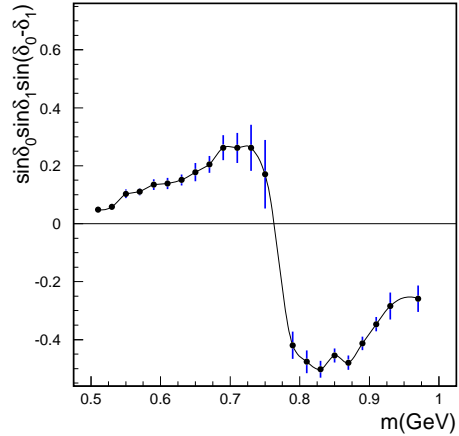}}
        \caption{{\sf
           The factor, $\sin\delta_0 \sin\delta_1
\sin(\delta_0-\delta_1)$, 
           as a function of the invariant mass $m$ of two-pion
system. 
           The data on $\pi\pi$ phase shifts are taken from
Ref.~\protect\cite{martin74}.
}}
        \label{fig2}
\end{minipage}
\end{figure}

Since we are only interested in a result at the leading twist, we follow the
helicity density matrix formalism developed in Refs.~\cite{jaffe95,jaffe96},
in which all spin dependence is summarized  in a {\it double} helicity
density matrix. We factor the process at hand 
into basic ingredients: the $N\rightarrow q$ distribution
function, the hard partonic 
$eq\rightarrow e'q'$ cross section,  the $q \rightarrow (\sigma, \rho)$
fragmentation, and the decay $(\sigma, \rho)\rightarrow \pi\pi$, all as
density matrices in helicity basis: 
\begin{eqnarray}
\left[{{d^6\sigma}\over{dx\, dy\, dz\, dm^2\, d\cos\Theta\,
d\phi}}\right]_{H'H}&&
\nonumber
\\
&&\hspace*{-2.8cm}={\cal F}\,_{H'H}^{h_1h'_1}\left[{{d^2\sigma(e
q\rightarrow e' q')}
\over{dx\, dy}}\right]_{h'_1h_1}^{h_2h'_2}
\left[{{d^2\hat{\cal M}}\over{dz\,
dm^2}}\right]_{h'_2h_2}^{H_1H'_1}
\left[{{d^2{\cal D}}\over{d\cos\Theta\, d\phi}}\right]_{H'_1H_1}\ ,
\label{hme} 
\end{eqnarray}
where $h_i(h_i')$ and $H(H')$ are indices labeling the helicity states of
quark and nucleon, and $H_1(H'_1)$ labeling the helicity state of resonance
the ($\sigma$, $\rho$). See Fig.~\ref{fig1}. In order
to incorporate the final state interaction,
we have separated the $q\rightarrow \pi^+\pi^-$ fragmentation process into 
two steps. First, the quark fragments into the resonance ($\sigma$,
$\rho$) ,  then the resonance decays into two pions, as shown at the top part
of the Fig.~\ref{fig1}.     

We first discuss two-meson fragmentation, first examined in
Ref.~\cite{collinsetal}. Here we introduce only those pieces  
necessary to describe $s$-$p$ interference in $\pi^+\pi^-$ production. 
A full account of these fragmentation functions will be given in 
Ref.~\cite{JJT2}. A two-meson fragmentation function can be defined 
by a natural generalization of the
single particle case.  Using the light-cone formalism of Collins and
Soper \cite{CollinsSoper}, the following replacement suffices,
\begin{equation}
  |hX\rangle_{\rm out}\:_{\rm out}\langle hX| \,\,\,\rightarrow\,\,\,
  |\pi^+\pi^-X\rangle_{\rm out}\:_{\rm out}\langle \pi^+\pi^-X|.
  \label{colsop}
\end{equation} 
The resulting two meson fragmentation function depends on the momentum
fraction of each pion, $z_1$, $z_2$, the $\pi\pi$ invariant mass, $m$, and
the angles $\Theta$ and $\phi$.  The subscript ``out'' places outgoing wave
boundary conditions on the $\pi\pi X$ state.  Two types of final state
interactions can generate a non-trivial phase: i) those between the two
pions, and ii) those between the pions and the hadronic state $X$.  We ignore
the latter because we expect the phase to average to zero when the sum on $X$
is performed ---
 $|\pi^+\pi^-X\rangle_{\rm out}\rightarrow |(\pi^+\pi^-)_{\rm out}X\rangle$.
Furthermore, if the two-pion system is well approximated by a single
resonance, then the resonance phase cancels in the product $|(\pi^+\pi^-)_{\rm
out}X\rangle\:\langle (\pi^+\pi^-)_{\rm out}X|$.  This leaves only
the interference between two partial waves as a potential source of an
asymmetry.  The final state phase of the two-pion system is determined by the
$\pi\pi\: {\cal T}$-matrix \cite{scattheory}.  We separate out the phase for
later consideration and analyze the (real) $\rho$-$\sigma$
interference fragmentation function as if the two particles were stable.

The $s$-$p$ interference fragmentation function
describes the emission of a $\rho (\sigma)$ with helicity $H_1$ from a quark 
of helicity
$h_2$, followed by absorption of $\sigma (\rho)$, with helicity $H'_1$ 
forming a quark of helicity $h'_2$. Conservation of angular momentum along
the $\hat{e}_3$ axis requires
\begin{equation}
H_1 + h_2' = H_1' + h_2\ .
\label{conservation}
\end{equation}
Parity and time reversal restrict the number of independent components  of
$\hat{\cal M}$:
\begin{eqnarray}
\hat{\cal M}^{sp(ps)}_{H_1'H_1,h_2h'_2} & = &
\hat{\cal M}^{sp(ps)}_{-H_1'-H_1,-h_2-h'_2}\quad {\rm 
	(parity)}\ ,\label{parity}\\
\hat{\cal M}^{sp}_{H_1'H_1,h_2h'_2}& = &
\hat{\cal M}^{ps}_{H_1H_1',h'_2h_2}\quad {\rm 
	(T-reversal)\ .}
\label{pt}
\end{eqnarray}
 Note that 
Eq.~(\ref{pt}) holds only after the T-reversal violating final state  
interaction between two pions is separated out.  After these symmetry 
restrictions, only two independent components remain, 
\begin{eqnarray}
\hat{\cal M}^{sp}_{00,++} & = &\hat{\cal M}^{ps}_{00,++} =
\hat{\cal M}^{sp}_{00,--}= \hat{\cal M}^{ps}_{00,--}\propto
\hat{q}_{_I}\ ,
\\
\hat{\cal M}^{sp}_{01,+-} & = &\hat{\cal M}^{ps}_{10,-+}
=\hat{\cal M}^{sp}_{0-1,-+} 
=\hat{\cal M}^{ps}_{-10,+-}\propto \delta \hat{q}_{_I}\ ,
\label{densityM}
\end{eqnarray}
and they can be identified with two novel interference fragmentation 
functions, $\hat q_{_I}$, $\delta \hat q_{_I}$,
where the subscript 
$I$ stands for interference. Here, to preserve clarity,  the flavor, $Q^2$, and $z$
have been suppressed.  The helicity
$\pm\frac{1}{2}$  states of quarks are denoted $\pm$, respectively. 
Hermiticity and time reversal invariance guarantee $\hat q_{_I}$
and $\delta \hat q_{_I}$ are real.

From Eq.~(\ref{densityM}) it is clear that the interference
fragmentation function, $\delta \hat{q}_{_I}$, is associated with quark 
helicity flip and is therefore chiral-odd. It is this feature
that enables us to access the chiral-odd quark transversity distribution 
in DIS.

Encoding this information into a double density matrix
notation, we define
\begin{eqnarray}
{d^2 \hat{\cal M}\over dz\, dm^2}
= &&
\Delta_0(m^2)\left\{I\otimes \bar\eta_0\,
\hat{q}_{_I}(z)
+\left(\sigma_+\otimes \bar\eta_-
+ \sigma_-\otimes \bar\eta_+\right)\delta\hat{q}_{_I}(z)\right\}
\Delta^*_1(m^2)
\nonumber
\\
& &
+\Delta_1(m^2)\left\{I\otimes \eta_0\,
\hat{q}_{_I}(z)
+\left(\sigma_-\otimes \eta_+ 
+ \sigma_+\otimes\eta_-\right)\delta\hat{q}_{_I}(z)
\right\}\Delta^*_0(m^2)\ ,
\label{fragmentation}
\end{eqnarray}
where $\sigma_\pm\equiv (\sigma_1\pm i\sigma_2)/2$ with
$\{\sigma_i\}$ the usual Pauli matrices. The $\eta$'s are $4\times 4$ matrices
in $(\sigma, \rho)$  helicity space with nonzero elements only in the first
column, and the $\bar\eta$'s are the transpose matrices 
 ($\bar\eta_0 = \eta_0^T, \bar\eta_+=\eta_-^T,
\bar\eta_-=\eta_+^T$), with the first rows $(0,0,1,0)$,
$(0,0,0,1)$, and $(0,1,0,0)$ for $\bar\eta_0$, $\bar\eta_+$, and $\bar\eta_-$,
respectively. The explicit definition
of the fragmentation functions will be given in Ref.~\cite{JJT2}.

The final state interactions between the two pions are  included
explicitly in
\begin{equation}
\Delta_0(m^2)=-i \sin\delta_0
e^{i\delta_0}\ ,\hspace*{1cm} 
\Delta_1(m^2)=-i \sin\delta_1
e^{i\delta_1}\ ,
\label{propagators}
\end{equation}
where $\delta_0$ and $\delta_1$ are the strong interaction $\pi\pi$ phase
shifts.
Here we have suppressed the $m^2$ dependence  of the phase shifts for simplicity. 
The decay process,
$(\sigma,\rho)\rightarrow \pi\pi$,  can be easily calculated and encoded into
the helicity matrix formalism. The result for the interference part is
\begin{equation}
{d^2 {\cal D}\over d\cos\Theta\, d\phi}
={\sqrt{6}\over 8\pi^2 m}
\, \sin\Theta \left[ie^{-i\phi}\left(\eta_--\bar\eta_-\right)
+ie^{i\phi}\left(\eta_+-\bar\eta_+\right)
-\sqrt{2}\cot\Theta \left(\bar\eta_0 +\eta_0\right)\right]\ .
\end{equation}
Here we have adopted the customary conventions for the $\rho$ polarization
vectors, $\vec{\epsilon}_{\pm} = {\mp}(\hat{e}_1\pm i\hat{e}_2)/\sqrt{2}$
and $\vec{\epsilon}_0=\hat{e}_3$ in its rest frame with $e$'s the unit vectors.

In the double density matrix notation, the quark distribution
function ${\cal F}$ can be expressed as \cite{jaffe96}
\begin{equation}
{\cal F} = {1\over 2} q(x)~I\otimes I + {1\over 2} \Delta
q(x)~\sigma_3 \otimes 
\sigma_3+{1\over 2}  \delta q(x)~
\left(\sigma_+\otimes\sigma_-+\sigma_-\otimes\sigma_+\right)\ ,
\label{calf}
\end{equation}
where the first matrix in the direct product is in the
nucleon helicity space and the second in the quark helicity space.
Here $q$, $\Delta q$, and $\delta q$ are the spin average, helicity
difference, and transversity distribution functions, respectively,
and their dependence on $Q^2$ has been suppressed.

The hard partonic process of interest here is essentially the  forward virtual
Compton scattering as
shown in the middle of Fig.~\ref{fig1}. For an {\it unpolarized} electron beam,
the resulting cross section is
\cite{jaffe96}
\begin{equation}
{d^2\sigma(e q\rightarrow e' q')\over dx\, dy}= 
\frac{e^4e_q^2}{8\pi Q^2}\left[\frac{1+(1-y)^2}{2y}
\left(I\otimes I + \sigma_3 \otimes 
\sigma_3\right)  +  { 2(1-y)\over y}
\left(\sigma_+\otimes\sigma_-+\sigma_-\otimes\sigma_+\right)\right]\ ,
\label{sigmahel}
\end{equation}
in the collinear approximation.  Here $e_q$ is the charge fraction 
carried by a quark. We have integrated out the azimuthal angle of 
the scattering plane.


Combining all the ingredients together, and integrating over $\Theta$ to 
eliminate the $\hat q_{_I}$ dependence, we obtain the transversity dependent
part of the cross section  for the production of two pions (kaons) in the
current fragmentation region for unpolarized electrons incident on  a
transversely polarized nucleon as follows
\begin{eqnarray}
{{d^5\sigma_\bot}\over{dx\, dy\, dz\, dm^2\,
d\phi}}
&=&-\frac{\pi}{2} \frac{e^4}{32\pi^3 Q^2 m}\, { 1-y\over y}
\sqrt{6}\, \cos{\phi}\,\,
\nonumber
\\*[7.2pt]
&&\times
\sin{\delta_0} \sin{\delta_1} 
\sin\left({\delta_0-\delta_1}\right)\,
\sum_a e_a^2 \delta q_a(x)\, \delta \hat{q}_{_I}^a(z)\ .
\label{cross-section}
\end{eqnarray}
Here the sum over $a$ covers all quark and antiquark flavors.

An asymmetry is obtained by dividing out the polarization independent
cross section,
\begin{eqnarray}
{\cal A}_{\bot\top}\equiv {d\sigma_\bot -d\sigma_\top
\over d\sigma_\bot +d\sigma_\top}
&=&-\frac{\pi}{4} {\sqrt{6}(1-y)\over 1+(1-y)^2}\, \cos{\phi}\,\,
\sin{\delta_0} \sin{\delta_1}
\sin\left(\delta_0-\delta_1\right)\,
\nonumber
\\*[7.2pt]
&&\hspace*{1cm}\times 
{\sum_a e_a^2 \delta q_a(x)\, \delta \hat{q}_{_I}^a(z)\over
\sum_a e^2_a q_a(x)
\left[ \sin^2\delta_0
\hat{q}_0^a(z)
+\sin^2\delta_1
\hat{q}_1^a(z)\right]}\ ,
\label{asymmetry}
\end{eqnarray}
where $\hat{q}_0$ and $\hat{q}_1$ are spin-average fragmentation functions
for the $\sigma$ and $\rho$ resonances, respectively. This asymmetry can be
measured either by flipping the target transverse spin or by binning events
according to the sign of the crucial azimuthal angle $\phi$. Note that this
asymmetry only requires a transversely  polarized nucleon target, but not a
polarized electron beam.

The flavor content of the asymmetry ${\cal A}_{\bot\top}$ can be
revealed by using isospin symmetry and charge conjugation
restrictions.  For $\pi^+\pi^-$ production, isospin symmetry gives
$\delta\hat{u}_{_I} = - \delta\hat{d}_{_I}$ and $\delta\hat{s}_{_I} =
0$.  Charge conjugation implies $\delta \hat{q}^a_{_I}=-\delta
\hat{\bar q}^a_{_I}$.  Thus there is only one independent interference
fragmentation function for $\pi^+\pi^-$ production, and it may be factored
out of the asymmetry, $\sum_a e_a^2\delta q_a\delta\hat{q}_{_I}^a=[{4/
9} (\delta u - \delta\bar u)- {1/ 9} (\delta d - \delta\bar
d)]\delta\hat{u}_{_I}$.   Similar application of isospin
symmetry and charge conjugation to the $\rho$ and $\sigma$
fragmentation functions that appear in the denominator of
Eq.~(\ref{asymmetry}) leads to a reduction in the number of
independent functions: $\hat{u}_i =\hat{d}_i=\hat{\bar u}_i=\hat{\bar
d}_i$ and $\hat{s}_i=\hat{\bar s}_i$ for $i=\{0,1\}$. 
For other systems the situation is more complicated due to the
relaxation of Bose symmetry restrictions. For example, for 
$K\overline K$ system, $\delta \hat{q}^a_{_I}=-\delta
\hat{\bar q}^a_{_I}$ still holds, but $\delta\hat{u}_{_I}$,
$\delta\hat{d}_{_I}$, and $\delta\hat{s}_{_I}$, are in general
independent. We also note
that application of the Schwartz inequality puts an upper bound on the
interference fragmentation function, $\delta\hat{q}_{_I}^2\leq
4\hat{q}_0\hat{q}_1/3$ for each flavor.

Finally, a few comments can be made about our results. First, the
final state phase generated by the $s$-$p$ interference is crucial to this 
analysis. If the data are not kept  differential in enough kinematic 
variables, the effect will almost certainly average  to zero. We are 
particularly concerned about the two-meson invariant mass, $m$, 
where we can see explicitly that the interference averages to zero over 
the $\rho$ as shown in Fig.~\ref{fig2}.
Second, the transversity distribution is multiplied by the fragmentation
function $\delta\hat{q}_{_I}$. Note that the transversity distribution 
{\it always} appears in a product of two soft QCD functions due to its 
chiral-odd nature. In order to disentangle the transversity
distribution from the asymmetry, one may invoke the process 
$e^+ e^-\rightarrow (\pi^+\pi^- X) (\pi^+\pi^- X)$ to measure 
$\delta\hat{q}_{_I}$, or use QCD inspired models to estimate it~\cite{JJT2}.

To summarize,  we have introduced twist-two interference quark fragmentation 
functions in helicity density matrix formalism and shown how the nucleon's 
transversity distribution can be probed through the final state interaction 
between two mesons ($\pi^+\pi^-$, $\pi K$ or $K\overline{K}$) produced in 
the current fragmentation region in deep inelastic scattering on a 
transversely polarized nucleon.  The technique developed in this Letter 
can  also be applied to other processes. Straightforward applications 
include the longitudinally polarized nucleon, 
and $e^+ e^-\rightarrow (\pi^+\pi^- X) (\pi^+\pi^- X)$.  A somewhat more
complicated extension can be made to two-meson production in single
polarized nucleon-nucleon collisions --- $p\vec p_\perp\rightarrow
\pi^+\pi^- X$, {\it etc.}  These applications will be presented in future
publications.

\vspace*{1cm}

We would like to thank John Collins and Xiangdong Ji for helpful
conversations relating to this subject.

\newpage

\end{document}